\documentclass[prb,twocolumn,aps]{revtex4}
\input{psfig}

\begin{document}

\title{Electrical quantum measurement of a two
level system at arbitrary voltage and temperature
}

\author{A. Shnirman$^{1,2}$, D. Mozyrsky$^{2}$, and I. Martin$^{2}$}

\affiliation{
$^1$Institut f\"ur Theoretische Festk\"orperphysik,
Universit\"at Karlsruhe, D-76128 Karlsruhe, Germany\\
$^2$Theoretical Division, Los Alamos National Laboratory,
Los Alamos, NM 87545, USA}

\begin{abstract}
We calculate the noise spectrum of the output signal of a
quantum detector during continuous measurement of a two-level
system (qubit). We generalize the previous results obtained for
the regime of high voltages (when $eV$ is much larger than the qubit's
energy level splitting $\Delta$) to the case of arbitrary voltages and
temperatures. When $V \sim \Delta$ the output spectrum is essentially
asymmetric in frequency, i.e., the output signal is no longer classical.
In the emission (negative frequency) part of the spectrum the
peak due to the qubit's coherent oscillations can be 8 times higher
that the background pedestal. For $V < \Delta$ and $T=0$ the coherent peaks
do not appear at all.
\end{abstract}
\maketitle

\section{Introduction}
\label{sec:Introduction}

The problem of quantum measurements has been around since
the early days of quantum mechanics. The recent upsurge in
the interest to the quantum computing made it necessary to
investigate the properties of the real physical systems
used as quantum detectors. Thus far the most developed
and successful mesoscopic detectors are the under-damped
SQUIDs (or current biased Josephson junctions) which
perform switching (threshold, latching) measurements.
These systems were investigated for many years. Recently
such measurements were used to confirm the
coherent dynamics and manipulations of the superconducting qubits
\cite{Saclay_Manipulation_Science,Han_Manipulation_Science,%
Martinis_Rabi_PRL,Delft_Rabi}.
Other measuring devices operate in a smoother mode
similar to a (linear) amplifier. Especially interesting are electro-meters
whose conductance depends on the charge state of a nearby qubit.
Two families of electro-meters working in this regime have
mostly been discussed in the literature. These are the single
electron (Cooper pair) transistors (SET)
\cite{Our_PRB,Devoret_Schoelkopf_Nature,Averin_SET_Cotunneling,Our_RMP,%
Maassen_SET_Cotunneling,Johansson_Vertex_Corrections,Clerk_Girvin_PI,%
Johansson_PI} and the quantum point contacts (QPC)
\cite{Aleiner,Gurvitz,Levinson,Korotkov_Continuous,%
Buettiker_Martin,Averin_Korotkov,Goan_Trajectory,Goan_Dynamics,%
Pilgram_Buettiker,Clerk_Efficiency}.
Some of these schemes have been
implemented experimentally and used for quantum measurements
\cite{Buks,Schoelkopf_RFSET,Nakamura_Nature,Sprinzak,%
Aassime_RFSET,Chalmers_Oscillations}.

In general one distinguishes between strong and weak quantum
measurements. In the strong (von Neumann) measurement the meter
discriminates between the states of a qubit on a time scale much
shorter than the other time scales of the system, e.g., the
inverse level splitting of the qubit. Thus the qubit's internal
dynamics is irrelevant and the meter determines the basis in which
the measurement takes place (the eigenbasis of the measured
observable). Such measurements strongly resemble projection of the
qubit's state even though everything can be described by treating
the coupled system of the qubit and the meter quantum
mechanically. In the opposite, weak measurement limit, the
measurement time is relatively long and the state of the qubit may
in principle change during the measurement. Yet, there is a
special regime in which the measurements resemble projection. This
is the so called quantum non-demolition (QND) limit. For example,
if the coupling is longitudinal, i.e., if a spin (qubit) is placed
in a magnetic field along the $z$-axis, and the detector is
measuring the $\sigma_z$ observable of the spin, and there is no
extra environment capable of flipping the spin, then $\sigma_z$ is
conserved (non-demolition) and can be measured even if it takes a
long time. In all other weak coupling regimes the meter cannot
extract precise information about the initial state of the qubit.
One can only talk about continuous monitoring of the qubit by the
meter in which they influence each other. Studying such continuous
monitoring, e.g., in the stationary state, is still useful, as one
can extract physical characteristics of the meter and of the qubit
and later use them for manipulations, projection-like measurements
or quantum feedback control of the qubits \cite{Ruskov_Feedback}.

Continuous monitoring in the regime close to QND was considered in
Ref.~\cite{Our_Current_PRL}. The small deviation from the QND
limit causes rare spin flips. Thus the current (output signal) in
the meter shows the ``telegraph noise'' behavior. In the noise
spectrum of the current this translates into a Lorentzian peak
around zero frequency. Recently the QND measurements have 
been considered in the rotating frame of a spin (qubit) 
\cite{Averin_QND}. 
The regime far from QND was the main focus
of Refs.~\cite{Averin_Korotkov,Korotkov_Osc}. This regime is
realized, e.g., in the case of the transverse coupling between the
qubit and the meter (magnetic field along the z-axis while
$\sigma_x$ is being measured). The possibility to ``observe'' the
coherent oscillations of a qubit was analyzed. While, due to the
unavoidable noise, the oscillations can not be seen directly in
the output current, the spectral density of the current noise has
a peak at the frequency of the oscillations (Larmor frequency,
level splitting of the qubit). The laws of quantum mechanics limit
the possible height of the peak. In the case of a 100\% efficient
(quantum limited) detector the peak can be only 4 times higher
than the background noise pedestal \cite{Averin_Korotkov}.
Inefficiency of the detector reduces the height of the peak
further. Such inefficiency implies
\cite{Our_PRB,Devoret_Schoelkopf_Nature,Clerk_Efficiency} that
when the detector is used in the QND regime, the measurement time,
i.e., the time needed to discriminate between the states of the
qubit is longer than the lowest possible limit for this time,
i.e., the dephasing time. This, in turn, means that the meter
produces more noise than it is necessary for the measurement, or,
in other words, that some information obtained by the meter is not
transferred to the output signal.

All the results described above were obtained in the limit
when the voltage applied to the measuring device is much higher
than the qubit's energy level splitting $eV\gg \Delta$.
In particular, in this regime, the ``telegraph noise'' peak around
$\omega=0$ is absent in the case of the purely transverse
coupling, while in the intermediate regime (between longitudinal
and transverse) the two peaks coexist.
The output noise spectrum, in the regime $eV\gg \Delta$ is almost
symmetric (classical) at frequencies of
order and smaller than $\Delta$.
In this paper we relax the condition $eV\gg \Delta$.
We calculate the non-symmetrized current-current correlator in the
case of the purely transverse coupling between the
qubit and the meter for arbitrary voltage and temperature.
At low voltages, $eV \sim \Delta$, the output noise is essentially
asymmetric, i.e., the output signal is quantum. In other words,
we have to differentiate between the absorption ($\omega >0$) and the 
emission ($\omega<0$) spectra of the detector 
(see, e.g. Ref.~\cite{Gardiner_book}).
Thus, the detector
ceases to be a device able to translate quantum information into
the classical one, and the way the output is further measured becomes
important. The qubit produces two symmetrically
placed peaks in the output noise spectrum.
The ratio of the peaks' hight to the hight of the pedestal can
reach $8$ for the negative frequency peak. We also obtain a small
peak at $\omega=0$ (even for purely transverse coupling).

The rest of the paper is organized as follows. In the next section
we define the physical system and outline the computational
scheme.  In Section~\ref{sec:Formalism}, we derive general
expressions for average current and fluctuations spectrum, valid
for any voltage and temperature.  In section~\ref{sec:Results} we
provide results for the specific model of a qubit transversely
coupled to a quantum point contact. In the appendices we
establish connections and clarify distinctions between the
technique developed here and the other existing approaches. In
Appendix~\ref{app:MacDoland} we analyze the validity of the
MacDonald's formula for the current noise in our case. 
In Appendix~\ref{app:Beyond_BR} we analyze the applicability of 
the Bloch-Redfield approximation.
In Appendix~\ref{app:Majorana} we outline the perturbation theory
based on spin representation by Majorana fermions.

\section{The system}
\label{sec:System}

We study the quantum measurement process in which a quantum
point contact (QPC) is used as a measuring device. These 
devices are known to serve as effective 
meters of charge (see, e.g., Refs.~\cite{Field,Sprinzak_Charge,Buks,%
Kouwenhoven_Charge}). 
In this paper we consider the
simplest limit of a tunnel junction when the
transmissions of all the transport channels is much smaller
than unity and is controlled by the quantum state of a qubit.
This model has previously been used by many 
authors~\cite{Gurvitz,Korotkov_Continuous,Goan_Dynamics}.
The measuring properties of QPCs in a more general case  
of open channels have been
studied, e.g., in Refs.~\cite{Pilgram_Buettiker,Clerk_Efficiency}.

The tunnel junction limit is described
by the following Hamiltonian
\begin{eqnarray}
\label{eq:Hamiltonian}
H = \sum_{l} \epsilon_{l} c_{l}^{\dag}\,c_{l} +
    \sum_{r} \epsilon_{r} c_{r}^{\dag}\,c_{r}
\nonumber \\
+ H_{\rm sys} + \sum_{l,r} \Omega (c_{r}^{\dag}\,c_{l}\,e^{-ik} +
h.c.) \ ,
\end{eqnarray}
where the tunneling amplitude $\Omega$
may in principle depend on any operator of the measured
system (qubit). In this paper we focus on the case of
transverse coupling between the meter and the qubit
\begin{eqnarray}
H_{\rm
sys} &=& -(1/2)\Delta \sigma_z \\
\label{Eq:Omega}
\Omega &=& T_0 + T_1\sigma_x
\ ,
\end{eqnarray}
which is optimal for ``observation'' of the qubit's
coherent oscillations. The transmission amplitudes
$T_0$ and $T_1$ are assumed to be real positive (as
the phase of the transmission amplitude does not matter
in our case) and small (tunnel junction limit).
Obviously, we only have to consider $T_1<T_0$.
We do not, however, assume
$T_1 \ll T_0$, i.e., we allow for the detectors with large
(relative to the average output) response.
We have also introduced the counting operator: $e^{-ik} |m\rangle =
|m+1\rangle$, where $m$ is the number of electrons that have
tunneled trough the point contact. Using this trick we could in
principle study the full counting statistics of the
current~\cite{Levitov_Lee_Lesovik_Review}.

\begin{figure}
\centerline{\hbox{\psfig{figure=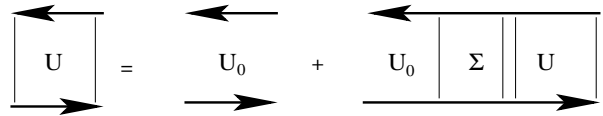,width=0.95\columnwidth}}}
\caption[]{\label{Figure:Dyson_Diagramm} Dyson equation.}
\end{figure}

Following Ref.~\cite{Schoeller_PRB} we integrate out the microscopic
degrees of freedom (the electrons) in the left and the right leads
and consider the time evolution of the reduced density matrix of
the system. This density matrix is a function of the measured
system's coordinates as well as of the variable $m$, i.e.,
$\hat\rho = \hat\rho(m_1,m_2)$. Due to the translational
invariance with respect to $m$ it is convenient to perform the
Fourier transform $\hat\rho(k_1,k_2) \equiv \sum_{m_1,m_2}
\hat\rho(m_1,m_2)e^{-ik_1 m_1+ik_2 m_2}$. In this representation
the operators $e^{\pm ik}$ in Eq.~(\ref{eq:Hamiltonian}) are
diagonal.
The master equation for the density matrix with the information
about the number of electrons that have tunneled was used in
Ref.~\cite{Gurvitz}. Here we do the same in the Fourier space
($k_1$ and $k_2$ indexes).
We write down the Dyson equation for the propagator of
the density matrix (see Fig.~\ref{Figure:Dyson_Diagramm}). Taking
the time derivative, one arrives \cite{Schoeller_PRB} at the
generalized master equation
\begin{equation}
\label{eq:Dyson_Equation}
\frac{d}{dt}\hat\rho(t)-
L_0 \hat\rho(t)
=
\int\limits_{t_0}^t dt'\; \Sigma(t-t')\;\hat\rho(t')
\ ,
\end{equation}
with the zeroth order Liouvillian
\begin{equation}
\label{eq:Zeroth_Liouvillian}
L_0 \equiv \left[ 1 \otimes i H_0^{\rm T} - i H_0 \otimes 1 \right]
\ .
\end{equation}
Throughout the paper we assume $\hbar=e=1$.
As we have integrated over the electronic degrees of freedom, the
Liouvillian $L_0$ as well as the Hamiltonian $H_0$ operate in the
direct product of the spin's ($|\uparrow/\downarrow\rangle$)
and the counting ($|m\rangle$) Hilbert spaces. As in the counting
space the zeroth Hamiltonian is zero (the number $m$ changes only
due to the tunneling) we have $H_0=H_{\rm sys}$
(mathematically rigorously one should write $H_0=H_{\rm sys}\otimes 0$).
We have chosen to present the Liouvillian as a super-operator
acting from the left on the density matrix regarded as a vector.
This way of writing makes the analysis easier and is especially
convenient for numerical simulations. As an example, we rewrite the
product $L_0 \hat\rho$ in the matrix form:
\begin{equation}
\label{Eq:Liouv_Matrix}
L_0\hat\rho=-\frac{i \Delta}{2}
\left(
\begin{array}{cccc}
\phantom{-}0 & \phantom{-}1 & -1 & \phantom{-}0\\
\phantom{-}1 & \phantom{-}0 & \phantom{-}0 & -1\\
-1 & \phantom{-}0 & \phantom{-}0 & \phantom{-}1\\
\phantom{-}0 & -1 & \phantom{-}1 & \phantom{-}0
\end{array}
\right)
\left(
\begin{array}{c}
\hat\rho_{11}\\
\hat\rho_{12}\\
\hat\rho_{21}\\
\hat\rho_{22}
\end{array}
\right)
\ .
\end{equation}

\begin{figure}
\centerline{\hbox{\psfig{figure=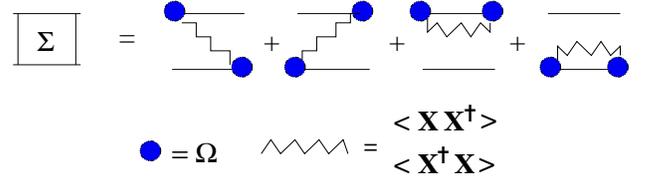,width=0.95\columnwidth}}}
\caption[]{\label{Figure:Sigma_Diagramm} The first order
approximation for the self-energy.}
\end{figure}

In the lowest non-vanishing approximation (second order in the tunneling
Hamiltonian) for the self-energy  we obtain
\begin{eqnarray}
\label{eq:Sigma_Born}
&&\Sigma(t) =
\nonumber \\
&&\phantom{+}\alpha_{+}^{*}(t)\,\Omega^{u}_{-k_1}\,U_0(t)\,\Omega^{d}_{k_2}
+ \alpha_{-}^{*}(t)\,\Omega^{u}_{k_1}\,U_0(t)\,\Omega^{d}_{-k_2}
\nonumber \\
&&+\alpha_{+}(t)\,\Omega^{d}_{k_2}\,U_0(t)\,\Omega^{u}_{-k_1}
+\alpha_{-}(t)\,\Omega^{d}_{-k_2}\,U_0(t)\,\Omega^{u}_{k_1}
\nonumber \\
&&-\alpha_{+}(t)\,\Omega^{u}_{k_1}\,U_0(t)\,\Omega^{u}_{-k_1}
-\alpha_{-}(t)\,\Omega^{u}_{-k_1}\,U_0(t)\,\Omega^{u}_{k_1}
\nonumber \\
&&-\alpha_{+}^{*}(t)\,\Omega^{d}_{-k_2}\,U_0(t)\,\Omega^{d}_{k_2} -
  \alpha_{-}^{*}(t)\,\Omega^{d}_{k_2}\,U_0(t)\,\Omega^{d}_{-k_2}
\ ,
\nonumber \\
\end{eqnarray}
where $\Omega^{u}_{\pm k_1} \equiv e^{\pm ik_1}\Omega^{u}$,
$\Omega^{d}_{\pm k_2} \equiv e^{\pm ik_2}\Omega^{d}$, $\Omega^{u}
\equiv (\Omega \otimes 1)$, $\Omega^{d} \equiv (1 \otimes
\Omega^{\rm T})$ and $U_0(t) \equiv e^{L_0 t}$. The superscripts
$u$ and $d$ stand for the "up" and "down" Keldysh contours. We
observe that all the matrix elements of the self-energy
$\Sigma(t)$ in Eq.~(\ref{eq:Sigma_Born}) are functions of $\Delta
k \equiv k_1-k_2$ only, i.e., they describe transitions which
conserve $m_1-m_2$. In particular they connect the diagonal
elements ($m_1=m_2$) with only the diagonal ones. Even
though we could have multiplied the factors $e^{\pm ik_{1/2}}$ in
Eq.~(\ref{eq:Sigma_Born}) and express $\Sigma$ as function of
$\Delta k$, we keep these factors separately in this particular
formula for future convenience. We also introduce the correlators
\begin{equation}
\alpha_{+}(t) \equiv \langle X(t)X^{\dag}(0)\rangle
\ ,
\end{equation}
and
\begin{equation}
\alpha_{-}(t) \equiv \langle X^{\dag}(t)X(0)\rangle
\ ,
\end{equation}
where
$X^{\dag}\equiv  \sum_{l,r} c_{r}^{\dag}\,c_{l}$.
Their Fourier transforms are:
\begin{equation}
\label{eq:alpha_plus}
\alpha_{+}(\omega) = \eta\,(\omega+V)
\left[\frac{1}{2}\coth\frac{\omega+V}{2T}+\frac{1}{2}\right]
\ ,
\end{equation}
and
\begin{equation}
\label{eq:alpha_minus}
\alpha_{-}(\omega) = \eta\,(\omega-V)
\left[\frac{1}{2}\coth\frac{\omega-V}{2T}+\frac{1}{2}\right]
\ ,
\end{equation}
where $\eta \equiv 2\pi\rho_{\rm L}\rho_{\rm R}$.
Fig.~\ref{Figure:Sigma_Diagramm} shows the diagrams that lead to
Eq.~(\ref{eq:Sigma_Born}).

Formally, Eq.~(\ref{eq:Dyson_Equation}) can be solved by
applying the Laplace transformation (that is, it becomes a simple
system of linear equations for each value of $s$):
\begin{equation}
\label{eq:Laplace_Solution} \hat\rho(k_1,k_2,s) = U(s,\Delta
k)\hat\rho_0 \ ,
\end{equation}
where
\begin{equation}
\label{Eq:U} U(s,\Delta k) \equiv (s-L_0-\Sigma(\Delta k,s))^{-1}
\ ,
\end{equation}
and $\hat\rho_0$ is the density matrix at $t=t_0$.

Further approximations are sometimes used to make the master
equation (\ref{eq:Dyson_Equation}) Markovian. When the dissipative
processes are slow in comparison with the unperturbed coherent
(Hamiltonian) dynamics, the Bloch-Redfield approximation is
appropriate. Within this approximation one substitutes
$\hat\rho(t') \rightarrow e^{-L_0(t-t')}\hat\rho(t)$ in the RHS of
Eq.~(\ref{eq:Dyson_Equation}). This leads to  $\Sigma(s)
\rightarrow \Sigma_{\rm BR}$ in Eq.~(\ref{Eq:U}), where
\begin{equation}
\label{Eq:Sigma_BR}
\Sigma_{\rm BR}
\equiv
\int_{0}^{\infty}dt\,\Sigma(t)e^{-L_0 t}
\ .
\end{equation}
In our case the validity domain of this approximation extends also
to the regime when the dissipative rates are bigger than $\Delta$,
i.e., when the qubit is over-damped. Below we will see that, due
to the smallness of $T_0$ and $T_1$, the  qubit can become
over-damped only at high voltages or temperatures, i.e., when
$V\gg\Delta$ or $T\gg\Delta$. In that case, however, the
self-energy $\Sigma(t-t')$ decays on the time scale of order $1/V$
or $1/T$ and, thus, is Markovian for slower processes. As we are
mostly interested in frequencies not much higher than $\Delta$, we
can still use $\Sigma_{\rm BR}$. In this paper we will mostly
employ the Bloch-Redfield approximation, which gives very simple
and transparent results. In Appendix~\ref{app:Beyond_BR}, however, 
we will use the non-Markovian expression (\ref{Eq:U}) to confirm that the
non-Markovian corrections are small.

As an example of the Bloch-Redfield approximation,
let us consider the regime $V\gg\Delta$ and $T=0$.
Then, for $|\omega|\ll V$ one has $\alpha_+(\omega)
= \eta(V+\omega)$, i.e., $\alpha_+(t) = \eta(V\delta(t) +
i\delta'(t))$, while $\alpha_-(\omega) = 0$, i.e., $\alpha_-(t) =
0$. Then using the definition (\ref{Eq:Sigma_BR}) we obtain
\begin{eqnarray}
\label{eq:Sigma_BR} \Sigma_{\rm BR}&=& \frac{\eta\,V}{2}
\left\{2e^{i\Delta
k}\Omega^{u}\,\Omega^{d}-\Omega^{u}\Omega^{u}-\Omega^{d}\Omega^{d}\right\}
\nonumber \\
&-&\frac{i\,\eta}{2} \left\{ e^{i\Delta
k}\Omega^{u}\,[L_0,\Omega^{d}]-e^{i\Delta
k}\Omega^{d}\,[L_0,\Omega^{u}] \right\}
\nonumber \\
&-&\frac{i\,\eta}{2} \left\{
\Omega^{u}\,[L_0,\Omega^{u}]-\Omega^{d}\,[L_0,\Omega^{d}] \right\}
\nonumber \\
&+& i\eta\delta_{\omega_c}(0) \left\{
\Omega^{u}\Omega^{u}-\Omega^{d}\Omega^{d} \right\} \ ,
\end{eqnarray}
where the ``value of the delta function at zero'',
$\delta_{\omega_c}(0)$, should be understood as a constant
depending on the high energy cut-off $\omega_c$ (diverging with
it). Note that $\Omega^{u}\,\Omega^{d} = \Omega^{d}\,\Omega^{u}$. The
last term of Eq.~(\ref{eq:Sigma_BR}) can be rewritten as
$i\eta\delta_{\omega_c}(0)\,[\Omega^2,...]$. Thus, this is a
renormalization of the spin's Hamiltonian. This renormalization is
usually disregarded either because it is small or, as in
Caldeira-Leggett's approach, since a counter term has been already
added in the initial Hamiltonian. Substituting the self-energy of
Eq.~(\ref{eq:Sigma_BR}) into Eq.~(\ref{eq:Dyson_Equation}) we
arrive at the generalized master equation obtained in
Ref.~\cite{Mozyrsky_Martin} by other technique.

\section{Calculation for any voltage and temperature}
\label{sec:Formalism}

We start by writing down the formally exact expression for the
correlator $\langle m(t)m(t')\rangle$ for $t > t'$
(see Fig.~\ref{Figure:mm_correlator}):
\begin{figure}
\centerline{\hbox{\psfig{figure=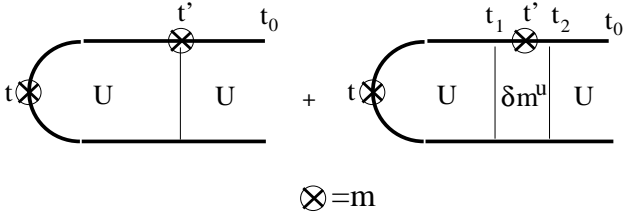,width=0.95\columnwidth}}}
\caption[]{\label{Figure:mm_correlator} Diagrammatic representation
of the correlator $\langle m(t)m(t')\rangle$.}
\end{figure}
\begin{eqnarray}
\label{eq:mm_exact} &&\langle \hat m(t)\hat m(t')\rangle = {\rm
Tr}\left[m^{u} U(t,t') m^{u}U(t',t_0)\hat\rho_0\right]
\nonumber \\
&&+\int\limits_{t'}^{t}dt_1 \int\limits_{t_0}^{t'}dt_2 {\rm
Tr}\left[m^{u} U(t,t_1) \delta
m^{u}(t_1,t',t_2)U(t_2,t_0)\hat\rho_0\right] \ ,
\nonumber \\
\end{eqnarray}
where $m^{u} \equiv (\hat m \otimes 1) = i\partial/\partial k_1$
is the bare vertex $\hat m$ on the upper Keldysh branch while
$\delta m^{u}(t_2,t',t_1)$ is the vertex correction. The
importance of the vertex corrections was recently pointed out in
Ref.~\cite{Johansson_Vertex_Corrections}. The ${\rm Tr[...]}$
operator is the trace of the density matrix (not of the
super-matrices like $\Sigma$). The trace over the $k_{1,2}$
indexes is calculated as $\int_{-\pi}^{\pi}\int_{-\pi}^{\pi}
\frac{dk_1 dk_2}{(2\pi)^2} \,2\pi\delta(k_1-k_2) ...$. For the
stationary state properties, which do not depend on the initial
density matrix, the trace operator over $k_{1,2}$ reduces to
taking the limit $\Delta k \rightarrow 0$. This follows from the
fact that all the propagators depend on $\Delta k$ only.

\begin{figure}
\centerline{\hbox{\psfig{figure=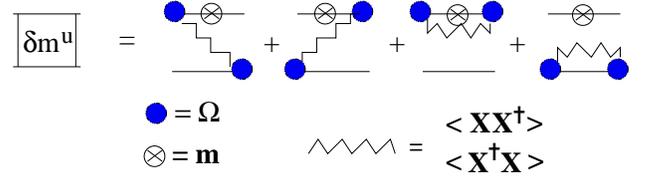,width=0.95\columnwidth}}}
\caption[]{\label{Figure:vertex_correction} The first order
approximation for the vertex correction.}
\end{figure}

For the vertex correction we use the same approximation we
have employed for the self-energy (\ref{eq:Sigma_Born})
(see Fig.~\ref{Figure:vertex_correction}).
One
easily obtains $\delta m^{u}(t_1,t',t_2)$ from
Eq.~(\ref{eq:Sigma_Born}) inserting the operator $m^{u}$ in all
the terms between the $\Omega$ operators (from either side of
$U_0$ as $m^{u}$ commutes with $U_0$):
\begin{eqnarray}
\label{eq:Vertex_Correction} &&\delta m^{u}(t_1,t',t_2)
\nonumber \\
&&=\alpha_{+}^{*}(t_1-t_2)\,\Omega^{u}_{-k_1}\,U_0(t_1-t_2)\,m^{u}\,
\Omega^{d}_{k_2}
+ ... \ ,
\end{eqnarray}
where $...$ stand for seven more terms obtained in the same way
from Eq.~(\ref{eq:Sigma_Born}). We observe that $\delta
m^{u}(t_1,t',t_2)=\delta m^{u}(t_1-t_2)$ in this approximation
does not explicitly depend on $t'$. We differentiate
Eq.~(\ref{eq:mm_exact}) over $t'$ and $t$ and use the master
equation (\ref{eq:Dyson_Equation}) written in the form $dU/dt =
(L_0+\Sigma)*U = U*(L_0+\Sigma)$ to obtain:
\begin{eqnarray}
\label{eq:II_m} &&\langle I(t)I(t') \rangle ={\rm Tr}\,\left[m^{u}
A_{(t-t')}U_{(t'-t_0)}\hat\rho_0\right] \nonumber
\\
&&+{\rm
Tr}\left[m^{u}\left\{(L_0+\Sigma)*U*A\right\}_{(t-t')}\,
U_{(t'-t_0)}\hat\rho_0\right]
\nonumber \\
&&-{\rm
Tr}\left[m^{u}\left\{(L_0+\Sigma)*U\right\}_{(t-t')}\,\left\{B*U
\right\}_{(t'-t_0)}\hat\rho_0\right]\ , \nonumber \\
\end{eqnarray}
where $A\equiv \delta m^{u} - \Sigma\, m^{u}$ and $B\equiv \delta
m^{u} - m^{u} \Sigma$. The convolutions are defined as
$\{g*f\}_{(t-t')} \equiv \int_{t'}^{t}dt_1 g(t-t_1)f(t_1-t')$ and,
analogously,
$\{g*f*h\}_{(t-t')} \equiv \int_{t'}^{t}dt_1 \int_{t'}^{t_1}dt_2
g(t-t_1)f(t_1-t_2)h(t_2-t')$. The symbol $L_0$ in
Eq.~(\ref{eq:II_m}) should be understood as a local in time
kernel, i.e., $L_0(t-t')=L_0 \delta(t-t'-0)$.
From
Eqs.~(\ref{eq:Sigma_Born}) and (\ref{eq:Vertex_Correction}) we
obtain
\begin{eqnarray}
\label{eq:A} &&A(t) =
\nonumber \\
&&\phantom{-}\alpha_{+}(t)\,\Omega^{d}_{k_2}\,U_0(t)\,\Omega^{u}_{-k_1}
- \alpha_{-}(t)\,\Omega^{d}_{-k_2}\,U_0(t)\,\Omega^{u}_{k_1}
\nonumber \\
&&-\alpha_{+}(t)\,\Omega^{u}_{k_1}\,U_0(t)\,\Omega^{u}_{-k_1}
+\alpha_{-}(t)\,\Omega^{u}_{-k_1}\,U_0(t)\,\Omega^{u}_{k_1} \
,
\nonumber \\
\end{eqnarray}
and
\begin{eqnarray}
\label{eq:B} &&B(t) =
\nonumber \\
&&-\alpha_{+}^{*}(t)\,\Omega^{u}_{-k_1}\,U_0(t)\,\Omega^{d}_{k_2}
+\alpha_{-}^{*}(t)\,\Omega^{u}_{k_1}\,U_0(t)\,\Omega^{d}_{-k_2}
\nonumber \\
&&-\alpha_{+}(t)\,\Omega^{u}_{k_1}\,U_0(t)\,\Omega^{u}_{-k_1}
+\alpha_{-}(t)\,\Omega^{u}_{-k_1}\,U_0(t)\,\Omega^{u}_{k_1} \
.
\nonumber \\
\end{eqnarray}
Next we note the following property of the super-operators $L_0$,
$\Sigma$, and $A$: ${\rm Tr}\left[L_0 ....\right]=0$, ${\rm
Tr}\left[\Sigma ....\right]=0$, and ${\rm Tr}\left[A
....\right]=0$. This allows us to simplify Eq.~(\ref{eq:II_m}):
\begin{eqnarray}
\label{eq:II_simple} &&\langle I(t)I(t') \rangle ={\rm Tr}\,\left[
[m^{u},A]_{(t-t')}U_{(t'-t_0)}\hat\rho_0\right] \nonumber
\\
&&+{\rm
Tr}\left[\left\{[m^{u},\Sigma]*U*A\right\}_{(t-t')}\,
U_{(t'-t_0)}\hat\rho_0\right]
\nonumber \\
&&-{\rm
Tr}\left[\left\{[m^{u},\Sigma]*U\right\}_{(t-t')}\,\left\{B*U
\right\}_{(t'-t_0)}\hat\rho_0\right]\ , \nonumber \\
\end{eqnarray}
and the commutators are readily calculated:
\begin{eqnarray}
\label{eq:[mA]} &&[m^{u},A]_{(t)} =
\nonumber \\
&&\phantom{+}\alpha_{+}(t)\,\Omega^{d}_{k_2}\,U_0(t)\,\Omega^{u}_{-k_1}
+ \alpha_{-}(t)\,\Omega^{d}_{-k_2}\,U_0(t)\,\Omega^{u}_{k_1} ,
\nonumber \\
\end{eqnarray}
\begin{eqnarray}
\label{eq:[mSigma]} &&[m^{u},\Sigma]_{(t)} =
\nonumber \\
&&\alpha_{+}^{*}(t)\,\Omega^{u}_{-k_1}\,U_0(t)\,\Omega^{d}_{k_2} -
\alpha_{-}^{*}(t)\,\Omega^{u}_{k_1}\,U_0(t)\,\Omega^{d}_{-k_2}
\nonumber \\
&&\alpha_{+}(t)\,\Omega^{d}_{k_2}\,U_0(t)\,\Omega^{u}_{-k_1}
-\alpha_{-}(t)\,\Omega^{d}_{-k_2}\,U_0(t)\,\Omega^{u}_{k_1} \
.
\nonumber \\
\end{eqnarray}
We observe that the super-operators $A$, $B$, $[m^{u},A]$, and
$[m^{u},\Sigma]$ do not contain the $\hat m$ operators, i.e.,
there are no differentiations over $k_{1,2}$ left in
Eq.~(\ref{eq:II_simple}). Thus we can safely perform the limit
$\Delta k \rightarrow 0$. We introduce the functions
$\alpha(t)\equiv \alpha_{+}(t)+\alpha_{-}(t)$ and $\beta(t)\equiv
\alpha_{+}(t)-\alpha_{-}(t)$ and obtain for $k_1=k_2$
\begin{eqnarray}
\label{eq:ABmAmSigma}
&&A(t)=-\beta(t)\Omega^{u}U_0(t)\Omega^{u}
+\beta(t)\Omega^{d}U_0(t)\Omega^{u}\ , \nonumber \\
&&B(t)=-\beta^{*}(t)\Omega^{u}U_0(t)\Omega^{d}
-\beta(t)\Omega^{u}U_0(t)\Omega^{u}\ , \nonumber \\
&&[m^{u},A]_{(t)}=\alpha(t)\Omega^{d}U_0(t)\Omega^{u}\ ,
\nonumber \\
&&[m^{u},\Sigma]_{(t)}=\beta^{*}(t)\Omega^{u}U_0(t)\Omega^{d} +
\beta(t)\Omega^{d}U_0(t)\Omega^{u} \ . \nonumber \\
\end{eqnarray}

Finally, for the stationary state we take $t_0 \rightarrow
-\infty$ and obtain (for $t>t'$)
\begin{eqnarray}
\label{eq:II_stationary} &&\langle I(t)I(t') \rangle ={\rm
Tr}\,\left[ [m^{u},A]_{(t-t')}\hat\rho_{\rm st}\right] \nonumber
\\
&&+{\rm
Tr}\left[\left\{[m^{u},\Sigma]*U*A\right\}_{(t-t')}\hat\rho_{\rm
st} \right]
\nonumber \\
&&-{\rm
Tr}\left[\left\{[m^{u},\Sigma]*U\right\}_{(t-t')}\,\left\{B*1
\right\}_{(\infty)}\hat\rho_{\rm st}\right]\ , \nonumber \\
\end{eqnarray}
where by definition $\left\{B*1 \right\}_{(\infty)} =
\int_0^{\infty}dt_1 B(t_1) = B(s=+0)$ and $B(s)$ is the Laplace 
transform of $B(t)$. The stationary
density matrix is given by
\begin{equation}
\label{Eq:Staitionary_rho} \hat\rho_{\rm st} = s\,U(s,\Delta
k)\hat\rho_0|_{s\rightarrow 0,\Delta k\rightarrow 0} \ .
\end{equation}

Analogously we find the
expression for the average current:
\begin{equation}
\label{eq:Current} \langle I(t) \rangle = {\rm
Tr}\left[\left\{[m^{u},\Sigma]*U\right\}_{(t-t_0)} \hat\rho_0
\right]\ ,
\end{equation}
which in the stationary regime becomes:
\begin{equation}
\label{eq:Stationary_Current} \langle I \rangle = {\rm
Tr}\left[\left\{[m^{u},\Sigma]*1\right\}_{\infty} \hat\rho_{\rm
st} \right] \ .
\end{equation}
Equations (\ref{eq:ABmAmSigma}), (\ref{eq:II_stationary}), and
(\ref{eq:Stationary_Current}) constitute the central result of
this chapter. They allow us to calculate the current-current
correlator and the average current in the first order approximation
for the self-energy and the vertex corrections.

\section{Average current and noise spectrum}
\label{sec:Results}

To formulate our results in a compact way it is convenient to
introduce the two following functions:
\begin{eqnarray}
&&s(\omega)\equiv\frac{\alpha(\omega)+\alpha(-\omega)}{2\eta}=
\nonumber \\
&&\frac{(V+\omega)}{2}\coth\frac{V+\omega}{2T}+
\frac{(V-\omega)}{2}\coth\frac{V-\omega}{2T}
\ ,
\nonumber \\
\end{eqnarray}
and
\begin{eqnarray}
&&a(\omega)\equiv\frac{\beta(\omega)-\beta(-\omega)}{2\eta}=
\nonumber \\
&&\frac{(V+\omega)}{2}\coth\frac{V+\omega}{2T}-
\frac{(V-\omega)}{2}\coth\frac{V-\omega}{2T}
\ .
\nonumber \\
\end{eqnarray}
One can easily check that
$\alpha(\omega) = \eta(s(\omega) + \omega)$, while
$\beta(\omega) = \eta(a(\omega) + V)$.

For the average current in the stationary regime we obtain the
following expression
\begin{equation}
\label{eq:Stationary_Current_Final_G}
\langle I \rangle =
g_0 V + g_1 V\left(1 - \frac{\Delta}{V}
\frac{a(\Delta)}{s(\Delta)}\right) \ ,
\end{equation}
where we have introduced the conductances
$g_0\equiv \eta T_0^2$ and $g_1\equiv \eta T_1^2$.
If $T=0$ the result simplifies.
For $V<\Delta$ we obtain $\langle I \rangle = g_0 V$, i.e, no
contribution of the qubit. For $V>\Delta$ we have
$\langle I \rangle = g_0 V + g_1 V (1 - \Delta^2/V^2)$.
Finally, for $V \gg \Delta$ we obtain
$\langle I \rangle \approx g_0 V + g_1 V =
(I_{\uparrow_{x}} + I_{\downarrow_{x}})/2$, where we have introduced
the values of the current corresponding to the two $x$-projections of the
qubit: $I_{\uparrow_{x}/\downarrow_{x}} = \eta(T_0\pm T_1)^2 V = V(g_0+g_1\pm
2\sqrt{g_0 g_1})$. 
This result becomes intuitively clear if one notes that the 
relevant frequency scale of the tunneling process is equal to 
$V$ while the fluctuations of spin's observable $\sigma_x$ have a 
characteristic frequency $\Delta$. In the regime $\max[V,T] \ll \Delta$
the spin is mostly in the ground state and 
the tunneling electrons ``see'' the quantum mechanical average 
value of $\sigma_x$,
i.e. zero. Therefore the system behaves as if there was no spin 
present, i.e., one should substitute in Eq.~(\ref{Eq:Omega})
$\Omega \rightarrow \langle \Omega \rangle = T_0$. 
In the opposite regime, $\min[V,T] \gg \Delta$, the spin is 
in the mixed state  and the electrons sometimes ``see'' the spin in the 
state $|\uparrow_x\rangle$ and sometimes in the state 
$|\downarrow_x\rangle$. Thus the current is the average 
of $I_{\uparrow_{x}/\downarrow_{x}}$.

To find the output noise spectrum
we perform the Laplace transform of Eqs.~(\ref{eq:II_stationary})
\begin{eqnarray}
\label{eq:II_Laplace} &&\langle I^2_{s} \rangle ={\rm
Tr}\,\left[ [m^{u},A]_{(s)}\hat\rho_{\rm st}\right] \nonumber
\\
&&+{\rm
Tr}\left[\left\{[m^{u},\Sigma]_{(s)}U(s)A(s)\right\}\hat\rho_{\rm
st} \right]
\nonumber \\
&&-{\rm
Tr}\left[\left\{[m^{u},\Sigma]_{(s)}U(s)B(+0)
\right\}\hat\rho_{\rm st}\right]
\ ,
\end{eqnarray}
and, then, find the Fourier transform of the current-current correlator.
This last step is done using {\it Mathematica} as the expressions are
quite extended. The Laplace transforms of the correlators $\alpha(t)$
and $\beta(t)$ contain, as usual, the real and the imaginary parts.
It is possible to show that the imaginary part of $\alpha$ gives rise to the
renormalization of the system parameters, e.g., the Lamb shift of the
qubit's level splitting $\Delta$, while the imaginary part of $\beta$ 
is only important at very high frequencies and it ensures the 
causality of the meter's response functions (see Appendix \ref{app:Majorana}).
In what follows we neglect the Lamb shift as it is small compared to 
$\Delta$.
The full expression splits into three parts
$\langle I^2_{\omega}\rangle = C_1 + C_2 + C_3$.

The first term of Eq.~(\ref{eq:II_Laplace}) does not contain the 
evolution operator $U(s)$. It is, thus, expected to give a non-resonant 
contribution to the current-current correlator, i.e., 
the pedestal (shot noise):
\begin{eqnarray}
\label{Eq:Shot_Noise_omega}
&&C_1=
g_0\left(s(\omega)+\omega\right)
\nonumber \\
&&+
g_1\frac{s(\omega+\Delta)+s(\omega-\Delta)+2\omega}{2}
\nonumber \\
&&-g_1\frac{\Delta\left(2\Delta+s(\omega+\Delta)-s(\omega-\Delta)\right)}
{2s(\Delta)} \ . \nonumber \\
\end{eqnarray}
For the symmetrized noise power $S_I^{\rm shot}(\omega) \equiv 
C_1(\omega) + C_1(-\omega)$ we then obtain
\begin{equation}
\label{Eq:Shot_Noise_omega=0} S_I^{\rm shot}(\omega=0) = 2g_0
s(0)+2g_1 s(\Delta)\left(1-\frac{\Delta^2}{s^2(\Delta)}\right)
\ .
\end{equation}
The two terms in Eq.~(\ref{Eq:Shot_Noise_omega=0}) clearly
correspond to the two terms of
Eq.~(\ref{eq:Stationary_Current_Final_G}). This is a usual situation
for the shot noise. Yet, the Fano factors for these two
contributions are slightly different (at $T=0$ both are equal
$1$) .

The last two terms of Eq.~(\ref{eq:II_Laplace}) contain the
evolution operator $U(s)$ and, thus, are expected to produce 
resonant contributions. 
We, first, employ the Bloch-Redfield
approximation, i.e., we substitute the self-energy $\Sigma(s)$ in
Eq.~(\ref{Eq:U}) by $\Sigma_{\rm BR}$. Then we obtain the two
remaining contributions $C_2$ and $C_3$. The contribution $C_2$ is
\begin{eqnarray}
\label{Eq:peak_at_Delta}
C_2(\omega)&=&
\frac{(\delta I)^2\Gamma\Delta^2}
{(\omega^2-\Delta^2)^2 +4\Gamma^2\omega^2}
\times
\nonumber\\
&&\left(1-\frac{\Delta a(\Delta)+\omega a(\omega)}
{2V s(\Delta)} \right) \ ,
\end{eqnarray}
where
\begin{equation}
\label{Eq:Gamma}
\Gamma \equiv
g_1 s(\Delta)=
\frac{(\sqrt{I_{\uparrow_{x}}}-\sqrt{I_{\downarrow_{x}}})^2}{4V}
\,s(\Delta)
\end{equation}
is the qubit's dephasing rate, while 
$\delta I \equiv I_{\uparrow_{x}} -
I_{\downarrow_{x}} = 4V\sqrt{g_0 g_1}$
is the sensitivity of the
meter.

In the regime $\Gamma\ll\Delta$ we
obtain two peaks placed around $\omega=\pm\Delta$. This is how the
qubit's damped coherent oscillations are reflected in the output
noise. Interestingly, this contribution is symmetric, even though
the symmetry should not have been expected in general. An asymmetric
contribution would correspond to a change in the current-current 
susceptibility (finite frequency differential conductance) due to the 
presence of the spin. Such corrections have usually the Fano 
shape. As shown in Appendix~\ref{app:Majorana} the lowest 
order Fano resonances vanish in our model due to the non-universal
behavior of the QPC's response functions.

The contribution $C_2$ (Eq.~(\ref{Eq:peak_at_Delta}))
is a product of a Lorentzian and a reduction factor
in the brackets. The Lorentzian 
coincides with the one obtained in Ref.~\cite{Ruskov_Korotkov}. 
The reduction factor simplifies for $T=0$.
Then, if $V>\Delta$, it is given by
$(1-\Delta^2/V^2)$, while for $V<\Delta$ it is equal to $0$. In
the last case the measuring device can not provide enough energy
to excite the qubit and, therefore, the qubit remains in the
ground state and does not produce any additional noise. The ratio
between the peak's hight and the pedestal's hight is different for
positive and negative frequencies. In the limit $g_1\ll g_0$,
$T=0$, and $\Delta < V$ we obtain 
$C_1(\pm \Delta) \approx g_0 V (1 \pm \Delta/V)$ and 
$C_2(\pm \Delta) \approx 4g_0 V (1-\Delta^2/V^2)$ and, thus, 
\begin{equation}
\frac{C_2(\omega=\pm\Delta)}{C_1(\omega=\pm\Delta)}
\approx 4(1\mp\frac{\Delta}{V}) \ .
\end{equation}
For $\Delta \rightarrow V$ the ratio for the negative frequency peak
reaches $8$. In this limit, however, the peak's hight is zero.
For symmetrized spectra the maximal possible ratio is
$4$ (Ref.~\cite{Korotkov_Osc}). 
We see that the enhancement of the signal-to-noise ratio 
is due to the fact that the suppression of the
qubit's contribution to the noise by factor $(1-\Delta^2/V^2)$
is smaller than the suppression of the negative frequency background 
noise by factor $(1-\Delta/V)$.
An interesting question is what
exactly is observed in the experiments. If the setup for the
measurement of the noise would be absolutely passive,
like the photon counters in the fluorescence experiments,
it could measure only what the system emits, i.e. the noise
at negative frequencies \cite{Gavish_Levinson_Imry,Lesovik_Loosen}.
Moreover, if one is only interested in the {\it excess} 
noise, i.e., in the nonequilibrium addition to the noise power 
due to the finite transport voltage, then even an active detector
may be useful. Namely, as shown in 
Ref.~\cite{Gavish_Imry_Levinson_Yurke}, if the excess noise 
power is (almost) symmetric, it can be effectively measured 
by a finite temperature LCR filter. In our case the excess
noise consists of the shot noise, $C_1-C_1(V=0)$, and 
the coherent peaks $C_2$. While the second contribution is  
symmetric, the first one is only approximately symmetric 
in the limit $g_1 \ll g_0$. Thus, in this limit, the combination 
$C_2+C_1-C_1(V=0)$ can be measured. The question of what 
can be measured in the regime $g_1 \sim g_0$,
when the excess noise is essentially asymmetric, will be 
considered elsewhere.

Note also that far from the resonance, for $|\omega| \gg V$, the
reduction factor in Eq.~(\ref{Eq:peak_at_Delta}) 
becomes negative, creating a very small negative
contribution to the current-current correlator.
This is an artifact of the Bloch-Redfield approximation.
The non-Markovian corrections are expected to
compensate  this negative contribution so that the
correlator is positive and vanishes at high negative
frequencies.

As the voltage increases so that $\Gamma \sim \Delta$,
the peaks given by Eq.~(\ref{Eq:peak_at_Delta}) start to overlap.
Then they form a single peak around $\omega=0$ which starts getting
narrower. Finally, when $\Gamma \gg \Delta$, the width of the
peak scales as
$\Delta^2/\Gamma \ll \Gamma$. This is the strong measurement
or Zeno~\cite{Harris_Stodolsky} regime. The meter manages to
almost localize the qubit in one of the eigenstates of the
measured observable ($\sigma_x$). The rare transitions (flips)
give rise to the ``telegraph'' noise peak around $\omega=0$.
The stronger is the measurement ($\Gamma$) the longer is
the average time between the flips (Zeno effect) and, thus,
the narrower is the peak. In this regime the output is classical
and the reduction factor plays no role (is equal to unity).

Finally, for the last contribution $C_3$ we obtain
\begin{eqnarray}
&&C_3(\omega) =
\frac{\Gamma^3}{\omega^2+4\Gamma^2}\times\nonumber\\
&&\frac{\left[a(\Delta+\omega)+a(\Delta-\omega)\right]}{s^4(\Delta)}\times
\nonumber \\
&&\Big(4V\Delta s(\Delta)-2\Delta^2a(\Delta)
-
s^2(\Delta)\left[a(\Delta+\omega)+a(\Delta-\omega)\right]
\nonumber \\
&&-
\Delta s(\Delta)\left[a(\Delta+\omega)-a(\Delta-\omega)\right]\Big)
\ .
\end{eqnarray}
This contribution corresponds to a small peak around
$\omega=0$. For $\Gamma\ll\Delta$, $T=0$, $\omega\ll\Delta$,
and $V>\Delta$ we obtain
\begin{equation}
C_3(\omega)\approx \frac{4\Gamma^3}{\omega^2+4\Gamma^2}\,
\frac{\Delta^2}{V^2}\left(1-\frac{\Delta^2}{V^2}\right)
\ ,
\end{equation}
while for the same conditions but $V<\Delta$
the contribution $C_3$ vanishes.
To understand the physical meaning of the peak at zero
frequency we note, that due to the asymmetry of the correlators
$\alpha$ and $\beta$ the expectation values of the current
corresponding to the two eigenstates of the spin's
Hamiltonian $|\uparrow_z\rangle$ and $|\downarrow_z\rangle$
are different (classically they would be equal as in both
states $\langle \sigma_x \rangle=0$).
Indeed substituting into Eq.~(\ref{eq:Stationary_Current})
the density matrices $|\uparrow_z\rangle\langle\uparrow_z|$
or  $|\downarrow_z\rangle\langle\downarrow_z|$
instead of $\hat\rho_{\rm st}$, that is forcing the steady state
to be one of the eigenstates we obtain for the respective currents
$I=g_0 V + g_1(V\mp a(\Delta))$.
As the qubit is coherent (under-damped) the back-action noise
causes random transitions between the
eigenstates of the qubit's Hamiltonian.
This translates into the ``telegraph'' noise of
the current. The effect is governed by the ratio $\Delta/V$ and is
small in the limit when $V\gg\Delta$.
In the quantum Zeno regime ($\Gamma\gg\Delta$) the contribution
$C_3$ is always negligible as compared with $C_1$ and $C_2$.

\begin{figure}
\centerline{\hbox{\psfig{figure=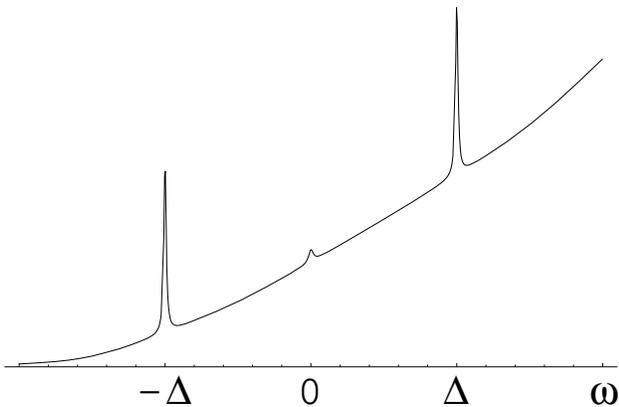,width=0.95\columnwidth}}}
\caption[]{\label{Figure:noise} The output noise for
the following parameters:
$V=1.5\Delta$, $T=0.1\Delta$, $g_0=0.01$, $g_1=0.0064$.}
\end{figure}

In Fig.~\ref{Figure:noise} we plot an example of the output noise
$\langle I^2_{\omega}\rangle$.

\section{Conclusions}
\label{sec:Conclusions}

We have calculated the output noise of the point contact used as a
quantum detector for arbitrary voltage and temperature. In the
regime $eV\sim \Delta$ and $T\ll \Delta$ the output noise is
essentially asymmetric. The qubit's oscillations produce two peaks
at $\omega =\pm \Delta$. The peaks have almost equal height and,
therefore, the negative frequency peak is much higher relative to
it's pedestal than the positive frequency one. The peak/pedestal
ratio can reach 8. As the negative frequencies correspond to
emission, this could be observed by further passive detectors. We
have also obtained a ``telegraph noise'' peak around $\omega=0$
for a purely transverse coupling. This peak appears due to the
quantum asymmetry of the noise spectra. It means that the detector
discriminates not only between the eigenstates of the measured
observable ($\sigma_x$ in our case) but also between the states of
different energy. The results of this paper are obtained for the
simplest and somewhat artificial model of a quantum detector. 
In particular, in this model, the leading contribution of the spin 
to the output current correlator is of the peak type at all voltages and 
vanishes at $V=0$. In general this should not be the case, 
as the coupling to an additional (discrete) degree of freedom usually 
changes the response functions of the continuum (Fano resonances).
In our system, however, certain properties of the response functions 
of the meter (see Appendix \ref{app:Majorana}) prevent the Fano 
resonances from appearing in the leading order of the 
perturbation expansion.
It would be interesting to perform analogous calculations for more
realistic detectors like SET's or QPC's with open channels.
 
Recently, Bulaevskii, Hru\^ska, and Ortiz\cite{Bulaevskii_Ortiz} 
studied the problem of a spin in a magnetic field interacting 
with tunneling electrons with arbitrary spin polarization.  
They considered the case of low dissipation, $\Gamma \ll \Delta$,
and the tunneling electrons were coupled to all the projections 
of the spin operator. 

\section{Acknowledgments}

We thank Yu.~Makhlin, G.~Sch\"on, D.~Averin, M.~B\"uttiker,
G.~Johansson, A. Rosch, L. Bulaevskii and Y. Levinson for fruitful
discussions. A.S. was supported by the EU IST Project SQUBIT, by
the DIP (Deutsch-Israelisches Projekt des BMBF), and by the CFN
(DFG). D.M. and I.M. were supported by the U.S. DOE.

\appendix
\section{Applicability of the MacDonald's formula}
\label{app:MacDoland}

For classical currents it is convenient to use the MacDonald's
formula to calculate the noise power of the current. Recently
this formula has been applied in Ref.~\cite{Ruskov_Korotkov}
to calculate the output noise of a QPC used as a measuring
device (the same system as in this paper). As only the limit
$V\gg\Delta$ was considered, the output signal was classical
and the calculation using the MacDonald's formula was well
justified. In this Appendix we clarify whether this approach is
applicable when the output is quantum.

The MacDonald's formula reads
\begin{equation}
S_I(\omega)=2\omega\int_{0}^{\infty}
\frac{d\langle m^2(t) \rangle}{dt} \sin(\omega t) dt
\ ,
\end{equation}
where $\sigma(t)=\langle m^2(t) \rangle$ is the dispersion of the
integral of current $m=\int_{0}^{t} I(t')dt'$.
One starts counting the charge that have tunneled starting from $t=0$.
One also assumes that at $t=0$ the
spin's density matrix is the stationary one (see
Eq.~(\ref{Eq:Staitionary_rho})) and that $m=0$ at $t=0$. Then,
since $\sigma(t=0) = 0$, we obtain
\begin{equation}
S_I(\omega)=-\omega^2\,\left[\sigma(s=i\omega+0)+\sigma(s=-i\omega+0)\right]
\ .
\end{equation}
To obtain $\sigma(s)$ we apply the propagator $U(s,\Delta k)$ (see
Eq.~(\ref{Eq:U})) to the stationary density matrix and, then apply
twice the operator $\hat m = i\partial/\partial \Delta k$ (as we
do it after the propagator, i.e. at the left-most end of the
Keldysh contour, we do not distinguish between $m^{u}$ and
$m^{d}$). As a result we obtain
\begin{equation}
\label{Eq:m_dispersion} \sigma(s) = -{\rm
Tr}\,\left[\frac{\partial^2}{\partial \Delta k^2} U(s,\Delta
k)\hat\rho_{\rm st}\right] \ .
\end{equation}
The derivative over $\Delta k$ in Eq.~(\ref{Eq:m_dispersion}) can
be calculated using Eq.~(\ref{Eq:U}):
\begin{equation}
\label{Eq:Ukk} \frac{\partial^2}{\partial \Delta k^2} U=
U\,\Sigma''\,U + 2\,U\,\Sigma'\,U\,\Sigma'\,U \ .
\end{equation}
From Eq.~(\ref{eq:Sigma_Born}) it is easy to obtain
$i\Sigma'=[(i\partial/\partial \Delta k)\Sigma]_{\Delta
k\rightarrow 0} = [m^{u},\Sigma]=A-B$ (see
Eq.~(\ref{eq:ABmAmSigma})) and $(i^2)\Sigma''=
[(i\partial/\partial \Delta k)^2 \Sigma]_{\Delta k\rightarrow 0} =
[m^{u},A]+h.c.$. After some algebra we conclude that the
MacDonald's formula gives for the noise $S_I$ an expression very
similar to the (symmetrized) Eq.~(\ref{eq:II_Laplace}). However,
while in the last line of Eq.~(\ref{eq:II_Laplace}) there is
$B(+0)$, the MacDonald's formula puts $B(s)$ into that place. With
this substitution we obtain for the $C_2$ contribution
\begin{eqnarray}
\label{Eq:C_2_MacDonald}
&&C_2(\omega)=
\frac{(\delta I)^2\Gamma\Delta^2}
{(\omega^2-\Delta^2)^2 +4\Gamma^2\omega^2}
\times\Big(1-
\nonumber\\
&&\frac{(\Delta+\omega) a(\Delta-\omega)+
                (\Delta-\omega) a(\Delta+\omega)+2\omega a(\omega)}
{4V s(\Delta)} \Big)\ .
\nonumber \\
\end{eqnarray}
In the classical limit $\Delta/V \rightarrow 0$ the result
(\ref{Eq:C_2_MacDonald}) coincides with the one obtained in full
quantum mechanical calculation (\ref{Eq:peak_at_Delta}). However
the corrections (even when $\Delta/V$ is small) are not
reproduced. Indeed at $T=0$ and $\Delta < V$ the reduction factor
in Eq.~(\ref{Eq:peak_at_Delta}) is $(1-\Delta^2/V^2)$ while
Eq.~(\ref{Eq:C_2_MacDonald}) gives $(1-\Delta^2/2V^2)$.

\section{Beyond the Bloch-Redfield approximation}
\label{app:Beyond_BR}

We have also calculated the output noise without using the
Bloch-Redfield approximation, i.e., substituting the non-Markovian
self energy $\Sigma(s)$ into Eq.~(\ref{Eq:U}). In the regime
$\Gamma \ll \Delta$ we found the following correction around
$\omega=\pm\Delta$
\begin{eqnarray}
\label{Eq:Non-Markovian}
 &&\delta C_2(\omega\approx\pm\Delta) \approx \frac{4
g_0 g_1 V \Delta (\omega^2-\Delta^2)}
{(\omega^2-\Delta^2)^2+8\Delta^2\Gamma^2}\times
\nonumber \\
&&\left(\frac{2(g_0-g_1)V\Delta + g_1 a(2\Delta)(s(\Delta)\mp\Delta)}
{s(\Delta)}\right) \
. \nonumber \\
\end{eqnarray}
This feature has a ``derivative'' (Fano) shape and it makes the
main peaks (Eq.~(\ref{Eq:peak_at_Delta})) a bit asymmetric. As far
as the coupling constants (conductances) are concerned, the
correction (\ref{Eq:Non-Markovian}) seems to be of the same order as the
terms $C_2$ and $C_3$. Thus, the question arises, what does the
Bloch-Redfield approximation exactly mean. Analyzing this question
deeper we note, that within this approximation the self-energy
$\Sigma$ and, consequently, the evolution operator $U$ factorize
into two parts: the part describing the diagonal (in the
eigen-basis of the qubit's Hamiltonian) elements of the density
matrix (two modes with eigenfrequencies around $\omega=0$) and the
part describing the off-diagonal elements (two modes with
eigenfrequencies around $\omega=\pm \Delta$). The first part is
responsible for the contribution $C_3$, while the second part in
responsible for $C_2$. The non-Markovian corrections to $\Sigma$
couple these two pairs of modes. These corrections are, however,
proportional to the deviations from the eigenfrequencies, e.g., to
$(\omega-\Delta)$ . Thus, they vanish exactly at the
eigenfrequencies. More rigorously, since the width of the
resonances is proportional to $\Gamma \sim g_1$,
the non-Markovian corrections carry
an additional factor of $g_0$ or $g_1$ within the resonances.
There, the Bloch-Redfield approximation is well justified and
the corrections are of the higher order in $g_0,g_1$.
Outside the resonances the Bloch-Redfield approximation may
be not justified. There, however, the
main contribution is the shot noise term $C_1$ which does not
depend on $U(s)$ and is not sensitive to the Bloch-Redfield
approximation. In Appendix~\ref{app:Majorana} we will show, 
that, indeed, the Fano shaped contribution (\ref{Eq:Non-Markovian})
is of the higher order than those, that could, in principle, have 
appeared together with the main peaks (\ref{Eq:peak_at_Delta}).

\section{Standard Keldysh calculation with Majorana fermions}
\label{app:Majorana}

It is possible to obtain Eqs.~(\ref{Eq:Shot_Noise_omega}) and
(\ref{Eq:peak_at_Delta})
using the standard Keldysh diagrammatic technique~\cite{Keldysh}.
For the two-level (spin-1/2) system it is convenient to
employ the mixed Dirac-Majorana-fermion representation
(see e.g. Ref~\cite{Tsvelik_Majorana}):
\begin{eqnarray}
\label{eq:Majorana_representation}
&&\sigma_{+} = \eta_z f\nonumber \\
&&\sigma_{-} = f^{\dag} \eta_z\nonumber \\
&&\sigma_{z} = 1-2f^{\dag}f
\ ,
\end{eqnarray}
where $f$ is the Dirac fermion while $\eta_z$ is the
Majorana fermion ($\eta_z = (g+g^{\dag})$, so that
$\{\eta_z,\eta_z\}=2$ ($g$ being another Dirac fermion)).

\begin{figure}
\centerline{\hbox{\psfig{figure=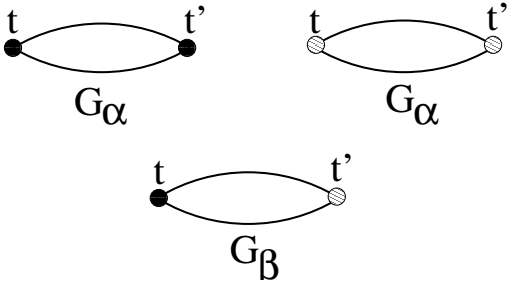,width=0.85\columnwidth}}}
\caption[]{\label{Figure:Loops} The graphical representation
of the Green's functions $G_{\alpha}$ (Eq.~(\ref{Eq:G_alpha})) and
$G_{\beta}$ (Eq.~(\ref{Eq:G_beta})). The loops are here to remind
that each of these Green's functions is actually a combination of
two electronic ones. In the approximations we use the electronic lines
appear only in such combinations.
}
\end{figure}

Our purpose is to calculate the correlator
$\langle I(t)I(t') \rangle$ which can be presented
as one of the components of the current-current Green's
function $G_{I}(t,t') = -i\langle T_{\rm K} I(t)I(t')\rangle$.
Namely
\begin{equation}
\label{Eq:II_G>}
\langle I(t)I(t') \rangle = i [\hat G_{I}]_{21} = i G_{I}^{>}
\ .
\end{equation}
In what follows we use the (Keldysh) notations explained in
Ref.~\cite{Rammer_Smith}. The current operator
is given by $I=i\Omega(X-X^{\dag})$, while the
tunneling Hamiltonian (the vertex of the perturbation
theory) is $H_{\rm T} = \Omega(X+X^{\dag})$.
It is, thus, convenient to introduce the two
following Green's functions:
\begin{eqnarray}
\label{Eq:G_alpha}
G_{\alpha} &\equiv& -i\langle T_{\rm K}
i\left[X(t)-X^{\dag}(t)\right]\cdot
i\left[X(t')-X^{\dag}(t')\right]\rangle
\nonumber \\
&=&-i\langle T_{\rm K}
\left[X(t)+X^{\dag}(t)\right]\cdot
\left[X(t')+X^{\dag}(t')\right]\rangle
\ ,\nonumber \\
\end{eqnarray}
and
\begin{equation}
\label{Eq:G_beta}
G_{\beta} \equiv -i\langle T_{\rm K}
i\left[X(t)-X^{\dag}(t)\right]\cdot
\left[X(t')+X^{\dag}(t')\right]\rangle
\ ,
\end{equation}
where the subscripts $\alpha$ and $\beta$ point
to an obvious relation to the functions $\alpha$
and $\beta$ introduced above. Indeed
$G_{\alpha}^{>} = -i\alpha(t-t')$ and
$G_{\alpha}^{<} = -i\alpha(t'-t)$,
while $G_{\beta}^{>} = \beta(t-t')$ and
$G_{\beta}^{<} = -\beta(t'-t)$.
The two lines of Eq.~(\ref{Eq:G_alpha}) might in principle
be different, for example, in the superconducting case.
In our case, however they are equal.
For these two Green's functions we use the
graphical representation shown in Fig.~\ref{Figure:Loops}.

Finally we introduce the fermionic Green's functions. For the
Majorana fermions we define $G_{\eta}\equiv -i\langle T_{\rm
K}\eta_z(t)\eta_z(t')\rangle$. It is easy to obtain the bare
Green's functions $G_{\eta,0}^{>}
= -i$ and $G_{\eta,0}^{<} = i$. For the $f$ fermions it
is convenient to use the Bogolubov-Nambu representation, i.e.,
$\Psi \equiv (f,f^{\dag})^{T}$ and $\Psi^{\dag}\equiv
(f^{\dag},f)$ and, then, $G_{\Psi}\equiv -i\langle T_{\rm
K}\Psi(t)\Psi^{\dag}(t')\rangle$.

The bosonic functions $G_{\alpha}$ and $G_{\beta}$
describe the reservoirs and, therefore, are well
approximated by their unperturbed values. Making 
this approximation we neglect a possibility of the 
spin-induced correlations in the reservoirs (leads),
i.e., the Kondo effect. This is justified if 
$\max{(T,V)} > T_{\rm K}$. 

The fermionic Green's function $G_{\Psi}$ and $G_{\eta}$ describe 
the spin, which can be driven far out of equilibrium.
Thus, we find these functions from the the kinetic (Dyson) equations.
\begin{equation}
\label{Eq:Dyson_Psi}
G_{\Psi}^{-1}=G_{\Psi,0}^{-1}-\Sigma_{\Psi}
\ ,
\end{equation}
\begin{equation}
\label{Eq:Dyson_eta}
G_{\eta}^{-1}=G_{\eta,0}^{-1}-\Sigma_{\eta}
\ ,
\end{equation}
where $\Sigma_{\Psi}$ and $\Sigma_{\eta}$ are the $f$ fermion's
and $\eta$ fermion's self-energies respectively. 
All the quantities in Eq.~(\ref{Eq:Dyson_Psi})
are matrices $4\times 4$ (in the Nambu and Keldysh spaces).
The operator $G_{\Psi,0}^{-1}$ is given by
\begin{equation}
\label{Eq:G_Psi_0}
G_{\Psi,0}^{-1}=
\left(
\begin{array}{cccc}
\omega - \Delta & 0 & 0 & 0\\
0 & \omega + \Delta & 0 & 0\\
0 & 0 & \omega - \Delta & 0\\
0 & 0 & 0 & \omega + \Delta
\end{array}
\right)
\ ,
\end{equation}
while for $G_{\eta,0}^{-1}$ we obtain
\begin{equation}
\label{Eq:G_eta_0}
G_{\eta,0}^{-1}=
\left(
\begin{array}{cc}
\omega/2 & 0 \\
0 & \omega/2 
\end{array}
\right)
\ .
\end{equation}
In Eqs.~(\ref{Eq:G_Psi_0}) and (\ref{Eq:G_eta_0})
we have neglected the infinitesimal terms responsible 
for, e.g., causality of the Green's functions. These are 
no longer needed when the finite self-energies are taken into 
account.

\begin{figure}
\centerline{\hbox{\psfig{figure=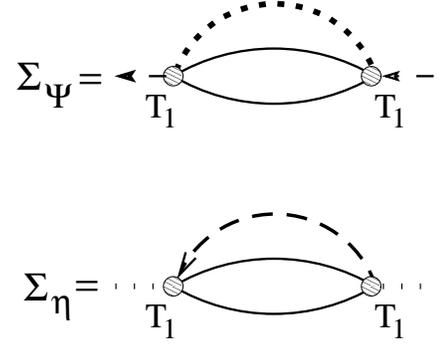,width=0.65\columnwidth}}}
\caption[]{\label{Figure:Sigma_ferm} The lowest order non-vanishing
diagrams for the self-energies $\Sigma_{\Psi}$ and $\Sigma_{\eta}$.
The solid lines are electronic
ones, the dashed line is the Green's function
of the Dirac ($f$) fermion and the dotted line
is the Majorana fermion's one.}
\end{figure}

For the self-energies we take the
lowest non-vanishing order approximation
shown in Fig.~\ref{Figure:Sigma_ferm}.
As usually, in the Keldysh space the self-energies
are presented as
\begin{equation}
\Sigma_{\Psi}=\left(\begin{array}{cc}
\Sigma_{\Psi}^{R} & \Sigma_{\Psi}^{K}\\
0 & \Sigma_{\Psi}^{A}
\end{array}\right)
\ ,\ \ \
\Sigma_{\eta}=\left(\begin{array}{cc}
\Sigma_{\eta}^{R} & \Sigma_{\eta}^{K}\\
0 & \Sigma_{\eta}^{A}
\end{array}\right)
\ .
\end{equation}
From Fig.~\ref{Figure:Sigma_ferm} it is easy to conclude
that $\Sigma_{\Psi}^{>} = i T_1^2\hat \lambda\,G_{\alpha}^{>}\,G_{\eta}^{>}$
and $\Sigma_{\Psi}^{<} = i T_1^2\hat \lambda\,G_{\alpha}^{<}\, G_{\eta}^{<}$,
where the Nambu matrix $\hat \lambda$ is defined
$\hat \lambda =
\left(
\begin{array}{cc}
\phantom{-}1 & -1\\
-1 & \phantom{-}1
\end{array}
\right)$. 
Analogously we obtain $\Sigma_{\eta}^{>}=iT_1^2
\,G_{\alpha}^{>}\,\left(\begin{array}{cc}1&-1\end{array}\right)
G_{\Psi}^{>}\left(\begin{array}{c}\phantom{-}1 \\ -1\end{array}\right)
$ and $\Sigma_{\eta}^{<}=iT_1^2
\,G_{\alpha}^{<}\,\left(\begin{array}{cc}1&-1\end{array}\right)
G_{\Psi}^{<}\left(\begin{array}{c}\phantom{-}1 \\ -1\end{array}\right)$.
To calculate the self-energies in the lowest order we can use the unperturbed 
retarded and advanced fermionic Green's functions. This is not so for the 
Keldysh component, which contains the information about the distribution function: 
$G_{\Psi/\eta}^{K}(\omega)=h_{\Psi/\eta}(\omega)
\left(G_{\Psi/\eta}^{R}(\omega)-G_{\Psi/\eta}^{A}(\omega)
\right)$. As was pointed out in Ref.~\cite{Parcollet_Hooley} the distribution 
functions $h_{\Psi/\eta}(\omega)$ are determined by the reservoirs even 
in the zeroth order. Thus they should be found self-consistently. 
After some algebra we find
\begin{equation}
\Sigma_{\Psi}^{R}-\Sigma_{\Psi}^{A}=
T_1^2\hat \lambda 
\left(G_{\alpha}^{K} + h_{\eta}(0)(G_{\alpha}^{R}-G_{\alpha}^{A})
\right)
\end{equation}
and 
\begin{equation}
\Sigma_{\Psi}^{K}=
T_1^2\hat \lambda 
\left((G_{\alpha}^{R}-G_{\alpha}^{A}) + h_{\eta}(0)G_{\alpha}^{K}
\right)
\ .
\end{equation}    
In what follows we will only need the Green's function $G_{\Psi}$.
Therefore, instead of proceeding with the self-consistent determination 
of the functions $h_{\Psi/\eta}$, we note that $h_{\eta}(0)=0$ just 
by symmetry (the self-consistent calculation gives the same).  
Thus we obtain
\begin{equation}
{\rm Im} \Sigma_{\Psi}^{R}(\omega) = -\hat\lambda \Gamma_{\omega}
\ ,
\end{equation}
where $\Gamma_{\omega}\equiv g_1 s(\omega)$ (see
Eq.~(\ref{Eq:Gamma}) where we have introduced
$\Gamma\equiv\Gamma_{\Delta}$). Analogously,
${\rm Im} \Sigma_{\Psi}^{A}(\omega) =
\hat\lambda \Gamma(\omega)$. The real parts of the retarded
and advanced self-energies give the non-equilibrium generalization
of the Lamb shift. Here we neglect it. For the
Keldysh component we have
\begin{equation}
\Sigma_{\Psi}^{K}(\omega)=-2 i g_1 \hat\lambda \omega
\ .
\end{equation}
Substituting the self-energy $\Sigma_{\Psi}$ into the 
Dyson equation (\ref{Eq:Dyson_Psi}) we obtain
\begin{eqnarray}
\label{Eq:G_Psi_(-1)}
&&G_{\Psi}^{-1}=G_{\Psi,0}^{-1}\nonumber \\
&&-\left(
\begin{array}{cccc}
-i\Gamma_{\omega} & i\Gamma_{\omega} & -2ig_1\omega & 2ig_1\omega\\
i\Gamma_{\omega} & -i\Gamma_{\omega} &  2ig_1\omega & -2ig_1\omega\\
0 & 0 & i\Gamma_{\omega} & -i\Gamma_{\omega}\\
0 & 0 & -i\Gamma_{\omega} & i\Gamma_{\omega}
\end{array}
\right)\ ,
\nonumber \\
\end{eqnarray}
which is easy to invert.
As a result we obtain
\begin{equation}
G_{\Psi}^{R/A}=\frac{\left(
\begin{array}{cc}
\omega+\Delta\pm i\Gamma_{\omega} & \pm i\Gamma_{\omega}\\
\pm i\Gamma_{\omega} & \omega-\Delta \pm i\Gamma_{\omega}
\end{array}
\right)}
{(\omega^2-\Delta^2) \pm 2i\omega\Gamma_{\omega}}
\ ,
\end{equation}
\begin{equation}
G_{\Psi}^{K}=\frac{2ig_1\omega\left(
\begin{array}{cc}
-(\omega+\Delta)^2 & \omega^2-\Delta^2\\
\omega^2-\Delta^2 & -(\omega-\Delta)^2
\end{array}
\right)}
{(\omega^2-\Delta^2)^2 + 4\omega^2\Gamma_{\omega}^2}
\ .
\end{equation}

\begin{figure}
\centerline{\hbox{\psfig{figure=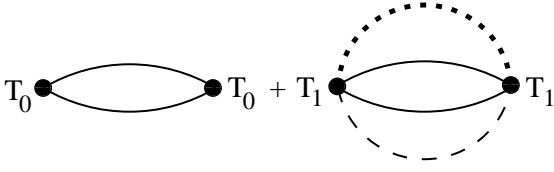,width=0.85\columnwidth}}}
\caption[]{\label{Figure:Keldysh_Shot} The diagrams leading to
Eq.~(\ref{Eq:Shot_Noise_omega}). The $f$-line should be found
from the Dyson (kinetic) equation as in Ref.~\cite{Parcollet_Hooley}.}
\end{figure}

Now we are ready to calculate the current-current
correlator (\ref{Eq:II_G>}). The lowest order diagrams
contributing to this correlator are shown in
Fig.~\ref{Figure:Keldysh_Shot}. They give
\begin{equation}
\label{Eq:Keldysh_Shot}
\delta G_{I}^{>}(t-t') = T_0^2 G_{\alpha}^{>}(t-t')+
iT_1^2 \Pi^{>}(t-t')G_{\alpha}^{>}(t-t')
\ ,
\end{equation}
where
\begin{eqnarray}
\Pi &\equiv& -i\langle T_{\rm K}\sigma_x(t)\sigma_x(t')\rangle
\nonumber \\
&=&-i\langle T_{\rm K}(\eta_z f + f^{\dag}\eta_z)_{(t)}
(\eta_z f + f^{\dag}\eta_z)_{(t')}
\rangle
\ .
\end{eqnarray}
One obtains Eq.~(\ref{Eq:Keldysh_Shot}) summing all possible
orientations of the lines in Fig.~\ref{Figure:Keldysh_Shot}.

In the Majorana representation the Green's function $\Pi$ is a
``two-particle'' Green's function (a bubble). To calculate it 
properly one has to take into account, e.g., the vertex corrections,
which seems to be complicated. Instead we use here an identity,
recently proven in Refs.~\cite{Coleman_Identity,Shnirman_Makhlin_Identity},
which reduces $\Pi$ to a single-fermion Green's function.
\begin{eqnarray}
\label{Eq:Bubble_to_Line_Reduction}
\langle \sigma_x(t)\sigma_x(t')\rangle=
\langle [f(t)+f^{\dag}(t)][f(t')+f^{\dag}(t')]
\rangle\ .
\end{eqnarray}

From Eq.~(\ref{Eq:Bubble_to_Line_Reduction})
it is easy to obtain the following relations
\begin{equation}
\Pi^{>}=\left(\begin{array}{cc}1&\ 1\end{array}\right)
G_{\Psi}^{>}\left(\begin{array}{c} 1 \\ 1\end{array}\right)
\ ,
\end{equation}
and
\begin{equation}
\Pi^{<}=-\left(\begin{array}{cc}1&\ 1\end{array}\right)
G_{\Psi}^{<}\left(\begin{array}{c} 1 \\ 1\end{array}\right)
\ .
\end{equation}

Performing usual Keldysh manipulations we obtain
\begin{equation}
\label{Eq:P_R-P_A}
\Pi^{R}-\Pi^{A} = \frac{-8i g_1 \omega\Delta^2}
{(\omega^2-\Delta^2)^2 + 4\omega^2\Gamma_{\omega}^2}
\ ,
\end{equation}
\begin{equation}
\label{Eq:P_K}
\Pi^{K} = \frac{-8i \Gamma_{\omega} \Delta^2}
{(\omega^2-\Delta^2)^2 + 4\omega^2\Gamma_{\omega}^2}
\ ,
\end{equation}
and, finally,
\begin{equation}
\label{Eq:Pi_>}
\Pi^{>} =\frac{1}{2}(\Pi^{K}+\Pi^{R}-\Pi^{A})=
\frac{-4i g_1(s(\omega)+\omega) \Delta^2}
{(\omega^2-\Delta^2)^2 + 4\omega^2\Gamma_{\omega}^2}
\ .
\end{equation}
We then obtain Eq.~(\ref{Eq:Shot_Noise_omega})
from Eqs.~(\ref{Eq:Keldysh_Shot}) and (\ref{Eq:Pi_>})
in the regime $\Gamma \ll \Delta$. In this case
we can approximate $\Pi^{>}$ by a sum of two delta
functions and perform the convolution in the second term
of Eq.~(\ref{Eq:Keldysh_Shot}). 

\begin{figure}
\centerline{\hbox{\psfig{figure=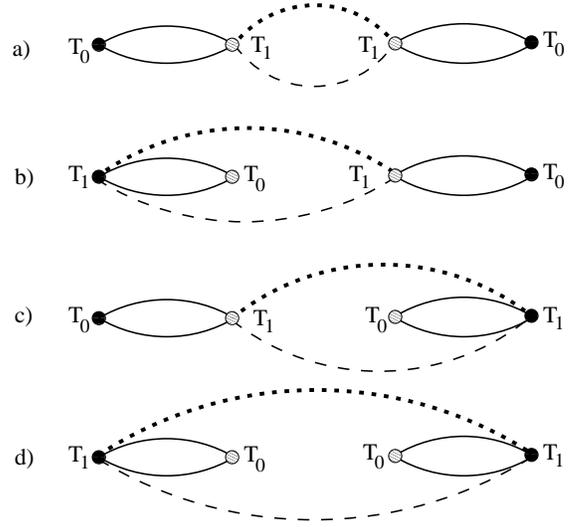,width=0.85\columnwidth}}}
\caption[]{\label{Figure:Keldysh_Peak} The diagrams leading to
Eq.~(\ref{Eq:peak_at_Delta}). The solid lines are electronic ones,
the dashed line is the Green's function of the Dirac ($f$) fermion
and the dotted line in the Majorana fermion's one.}
\end{figure}

The diagrams giving the peaks at $\omega=\pm \Delta$
are shown in Fig.~\ref{Figure:Keldysh_Peak}. They are
chosen out of many other second-order diagrams, since
only in these diagrams the spin's line (the combined
loop of the $f$ and Majorana fermions), which gives
the Green's function $\Pi$, carries the external
frequency $\omega$. In all other diagrams the spin's
lines participate in loops and are, thus, being integrated
over the frequency. Then the resonant structure is
washed out and one merely gets a second order correction
to the shot noise (pedestal). Although in 
Fig.~\ref{Figure:Keldysh_Peak} we draw the loops (bubbles) 
of $f$ and $\eta_z$ lines, we do not actually calculate 
those but use instead Eqs.~(\ref{Eq:P_R-P_A}) and (\ref{Eq:P_K}).

In it quite easy to calculate the first diagram
(Fig.~\ref{Figure:Keldysh_Peak}a). It is given
by $T_0^2 T_1^2 G_{\beta}(\omega)\Pi(\omega)[-G_{\beta}(\omega)]$.
In the other three diagrams the internal vertices,
over which the integration is performed, are not
further connected. Let us, for example consider
the diagram b). Acting, first, in the $(11,12,21,22)$
Keldysh coordinates (see Ref.~\cite{Rammer_Smith})
we see that the left electronic loop of this
diagram, after integration over the time of the
``free'' vertex, gives a ``Keldysh vector'' $\chi_j$,
where $j$ is the Keldysh index of the left external
vertex. Thus the whole diagram b) can be presented as
$T_0^2 T_1^2 \chi_j \hat D_{jk}$ (no summation
over $j$), where $D$ denotes the rest of the expression
which can be treated as a usual Keldysh matrix
($D$ in this case is given by $\Pi[-G_{\beta}]$ in
the $(R,A,K)$ coordinates). For $\chi_j$ we obtain
$\chi_j = [-G_{\beta}(0)]_{1j}-[-G_{\beta}(0)]_{2j} =
-G_{\beta}^{A}(0)$. Thus $\chi_j$ is actually a (Keldysh)
scalar and the diagram b) can be, finally, calculated as
$T_0^2 T_1^2 [-G_{\beta}^{A}(0)]\cdot \Pi(\omega)[-G_{\beta}(\omega)]$.
Collecting the rest of the diagrams we
obtain the following contribution to the current-current
Green's function
\begin{equation}
\delta G_I =T_0^2 T_1^2 [G_{\beta}(\omega)-G_{\beta}^{A}(0) \cdot \hat 1]
\,\Pi(\omega)\,[G_{\beta}^{R}(0) \cdot \hat 1 - G_{\beta}(\omega)]
\ .
\end{equation}

We calculate the Keldysh components of the Green's function
$G_{\beta}$ for $\omega \ll D$, where $D\propto \rho^{-1}_{\rm L/R}$ is 
the electronic bandwidth (the Fermi energy). As a result we obtain:
$G_{\beta}^{\rm R}(\omega) = \eta V[1 + iO(\omega/D)]$,
$G_{\beta}^{\rm A}(\omega) = -\eta V[1 - iO(\omega/D]$, and
$G_{\beta}^{\rm K}(\omega) = 2\eta a(\omega)$. The factors $1\pm
iO(\omega/D)$ are responsible for making the functions
$G_{\beta}^{\rm R}(t)$ and $G_{\beta}^{\rm A}(t)$ causal. As we
are interested in the low frequencies ($\omega \ll D$) 
we approximate those factors by $1$. Then we obtain
\begin{equation}
\label{Eq:G_I_Matrix}
\delta G_I = 4 g_0 g_1
\left(
\begin{array}{cc}
V & a(\omega)\\
0 & 0
\end{array}
\right)
\left(
\begin{array}{cc}
\Pi^{\rm R} & \Pi^{\rm K}\\
0 & \Pi^{\rm A}
\end{array}
\right)
\left(
\begin{array}{cc}
0 & -a(\omega)\\
0 & V
\end{array}
\right) \ ,
\end{equation}
and, finally, $\delta G^{\rm R/A}_I=0$ and
\begin{equation}
\label{Eq:G_I_K}
\delta G^{\rm K}_I = 2\delta G^{>}_I = 4 g_0 g_1 V^2
\left(\Pi^{\rm K} -\frac{a(\omega)}{V}(\Pi^{\rm R}-\Pi^{\rm A})
\right) \ .
\end{equation}
The first term in Eq.~(\ref{Eq:G_I_K}) is the standard 
contribution obtained in the high voltage limit, e.g., in 
Refs.~\cite{Averin_Korotkov,Korotkov_Osc}. It can also 
be obtained by treating the QPC as a linear amplifier, i.e.,
assuming the relation 
$I(t) = I_0(t) + 
\frac{\delta I}{2} \sigma_x (t)$~\cite{Averin_SQUID}. 
Then, the first term in Eq.~(\ref{Eq:G_I_K}) is the noise of 
the spin being amplified by the QPC. The interpretation of the 
second term in Eq.~(\ref{Eq:G_I_K}) is less trivial. 
Now these are the internal correlations of the QPC being amplified 
by the combined system of the spin and the QPC. This 
contribution is negligible when $\omega \ll V$ but 
is of the same order as the first one for $\omega \sim V$.
Using Eqs.~(\ref{Eq:P_R-P_A}) and (\ref{Eq:P_K})
we, finally, obtain the following contribution to the
current-current correlator
\begin{equation}
\label{Eq:peak_at_Delta_Keldysh}
i\delta G_I^{>}=
\frac{(\delta I)^2\Gamma_{\omega}\Delta^2}
{(\omega^2-\Delta^2)^2 +4\Gamma_{\omega}^2\omega^2}
\left(1-\frac{\omega a(\omega)}{V s(\omega)} \right) \ .
\end{equation}

The contribution (\ref{Eq:peak_at_Delta_Keldysh}) 
coincides with $C_2(\omega)$ (Eq.~(\ref{Eq:peak_at_Delta})) 
in the limit $\Gamma \ll \Delta$. 

Note that the response functions $G_{\beta}^{\rm R}(\omega)$ and 
$G_{\beta}^{\rm A}(\omega)$ vanish at $V=0$. This non-universal property
makes the spin's contribution to the equilibrium output current 
correlator to vanish. At any voltage and temperature, the contribution
(\ref{Eq:peak_at_Delta_Keldysh}) is of the ``peak'' type, and not of 
the Fano type. The Fano shaped resonances could have originated 
from the combination $\Pi^{\rm R} + \Pi^{\rm A}= 2{\rm Re}\,\Pi^{\rm R}$.  
Indeed, for $\Gamma\ll\Delta$ and 
$|\omega| \approx \Delta$ we obtain from Eq.~(\ref{Eq:P_R-P_A}) 
\begin{equation}
\label{Eq:P_R+P_A}
\Pi^{R}+\Pi^{A} \approx \frac{4\Delta^2}{s(\Delta)}\,\frac{\omega^2-\Delta^2}
{(\omega^2-\Delta^2)^2 + 4\omega^2\Gamma^2}
\ .
\end{equation}
This combination does not appear
in Eq.~(\ref{Eq:peak_at_Delta_Keldysh}) due to another non-universal 
property of the response functions $G_{\beta}^{\rm R}(\omega)$ and 
$G_{\beta}^{\rm A}(\omega)$ which are purely real up to the frequencies 
of order $D$. Had this not been the case, the Fano type contribution
would be similar to that of Eq.~(\ref{Eq:Non-Markovian}) but of 
a lower order in $g_0,g_1$ ($\propto g_0 g_1$).

\bibliographystyle{apsrev}
\bibliography{ref}
\end{document}